\documentclass[prl,twocolumn,showpacs,superscriptaddress,floatfix]{revtex4}

\usepackage{ifthen}
\usepackage{epsfig}
\usepackage{graphicx}
\usepackage{dcolumn}
\usepackage{mathrsfs}
\usepackage{subfigure}

\def\be{\begin{equation}}
\def\ee{\end{equation}}

\def\ptm{p_{T}}
\newcommand{\ppbar}{$p\bar{p}$}

\newcommand{\py}{{\sc pythia}}

\newcommand{\BRhgg}{$\cal{B}$($h \rightarrow \gamma \gamma$)}

\newcommand{\fb}{fb$^{-1}$}

\newcommand{\Mgg}{$m_{\gamma \gamma}$}
\newcommand{\Mh}{$m_{h}$}
\newcommand{\pt}{$p_{T}$}
\newcommand{\et}{$E_{T}$}
\newcommand{\ptgg}{$p_{T}^{~\gamma \gamma}$}
\newcommand{\gevcc}{GeV/$c^2$}
\newcommand{\gevc}{GeV/$c$}
\newcommand{\gev}{GeV}

\newcommand{\met}{$\not{\hspace{-.075in}E_{T}}$}

\begin{document}


%


\title{Search for a Fermiophobic Higgs Boson Decaying into Diphotons in \ppbar\ Collisions at $\sqrt{s} = 1.96$  TeV}

\affiliation{Institute of Physics, Academia Sinica, Taipei, Taiwan 11529, Republic of China} 
\affiliation{Argonne National Laboratory, Argonne, Illinois 60439} 
\affiliation{University of Athens, 157 71 Athens, Greece} 
\affiliation{Institut de Fisica d'Altes Energies, Universitat Autonoma de Barcelona, E-08193, Bellaterra (Barcelona), Spain} 
\affiliation{Baylor University, Waco, Texas  76798} 
\affiliation{Istituto Nazionale di Fisica Nucleare Bologna, $^v$University of Bologna, I-40127 Bologna, Italy} 
\affiliation{Brandeis University, Waltham, Massachusetts 02254} 
\affiliation{University of California, Davis, Davis, California  95616} 
\affiliation{University of California, Los Angeles, Los Angeles, California  90024} 
\affiliation{University of California, San Diego, La Jolla, California  92093} 
\affiliation{University of California, Santa Barbara, Santa Barbara, California 93106} 
\affiliation{Instituto de Fisica de Cantabria, CSIC-University of Cantabria, 39005 Santander, Spain} 
\affiliation{Carnegie Mellon University, Pittsburgh, PA  15213} 
\affiliation{Enrico Fermi Institute, University of Chicago, Chicago, Illinois 60637}
\affiliation{Comenius University, 842 48 Bratislava, Slovakia; Institute of Experimental Physics, 040 01 Kosice, Slovakia} 
\affiliation{Joint Institute for Nuclear Research, RU-141980 Dubna, Russia} 
\affiliation{Duke University, Durham, North Carolina  27708} 
\affiliation{Fermi National Accelerator Laboratory, Batavia, Illinois 60510} 
\affiliation{University of Florida, Gainesville, Florida  32611} 
\affiliation{Laboratori Nazionali di Frascati, Istituto Nazionale di Fisica Nucleare, I-00044 Frascati, Italy} 
\affiliation{University of Geneva, CH-1211 Geneva 4, Switzerland} 
\affiliation{Glasgow University, Glasgow G12 8QQ, United Kingdom} 
\affiliation{Harvard University, Cambridge, Massachusetts 02138} 
\affiliation{Division of High Energy Physics, Department of Physics, University of Helsinki and Helsinki Institute of Physics, FIN-00014, Helsinki, Finland} 
\affiliation{University of Illinois, Urbana, Illinois 61801} 
\affiliation{The Johns Hopkins University, Baltimore, Maryland 21218} 
\affiliation{Institut f\"{u}r Experimentelle Kernphysik, Universit\"{a}t Karlsruhe, 76128 Karlsruhe, Germany} 
\affiliation{Center for High Energy Physics: Kyungpook National University, Daegu 702-701, Korea; Seoul National University, Seoul 151-742, Korea; Sungkyunkwan University, Suwon 440-746, Korea; Korea Institute of Science and Technology Information, Daejeon, 305-806, Korea; Chonnam National University, Gwangju, 500-757, Korea} 
\affiliation{Ernest Orlando Lawrence Berkeley National Laboratory, Berkeley, California 94720} 
\affiliation{University of Liverpool, Liverpool L69 7ZE, United Kingdom} 
\affiliation{University College London, London WC1E 6BT, United Kingdom} 
\affiliation{Centro de Investigaciones Energeticas Medioambientales y Tecnologicas, E-28040 Madrid, Spain} 
\affiliation{Massachusetts Institute of Technology, Cambridge, Massachusetts  02139} 
\affiliation{Institute of Particle Physics: McGill University, Montr\'{e}al, Qu\'{e}bec, Canada H3A~2T8; Simon Fraser University, Burnaby, British Columbia, Canada V5A~1S6; University of Toronto, Toronto, Ontario, Canada M5S~1A7; and TRIUMF, Vancouver, British Columbia, Canada V6T~2A3} 
\affiliation{University of Michigan, Ann Arbor, Michigan 48109} 
\affiliation{Michigan State University, East Lansing, Michigan  48824}
\affiliation{Institution for Theoretical and Experimental Physics, ITEP, Moscow 117259, Russia} 
\affiliation{University of New Mexico, Albuquerque, New Mexico 87131} 
\affiliation{Northwestern University, Evanston, Illinois  60208} 
\affiliation{The Ohio State University, Columbus, Ohio  43210} 
\affiliation{Okayama University, Okayama 700-8530, Japan} 
\affiliation{Osaka City University, Osaka 588, Japan} 
\affiliation{University of Oxford, Oxford OX1 3RH, United Kingdom} 
\affiliation{Istituto Nazionale di Fisica Nucleare, Sezione di Padova-Trento, $^w$University of Padova, I-35131 Padova, Italy} 
\affiliation{LPNHE, Universite Pierre et Marie Curie/IN2P3-CNRS, UMR7585, Paris, F-75252 France} 
\affiliation{University of Pennsylvania, Philadelphia, Pennsylvania 19104}
\affiliation{Istituto Nazionale di Fisica Nucleare Pisa, $^x$University of Pisa, $^y$University of Siena and $^z$Scuola Normale Superiore, I-56127 Pisa, Italy} 
\affiliation{University of Pittsburgh, Pittsburgh, Pennsylvania 15260} 
\affiliation{Purdue University, West Lafayette, Indiana 47907} 
\affiliation{University of Rochester, Rochester, New York 14627} 
\affiliation{The Rockefeller University, New York, New York 10021} 
\affiliation{Istituto Nazionale di Fisica Nucleare, Sezione di Roma 1, $^{aa}$Sapienza Universit\`{a} di Roma, I-00185 Roma, Italy} 

\affiliation{Rutgers University, Piscataway, New Jersey 08855} 
\affiliation{Texas A\&M University, College Station, Texas 77843} 
\affiliation{Istituto Nazionale di Fisica Nucleare Trieste/Udine, $^{bb}$University of Trieste/Udine, Italy} 
\affiliation{University of Tsukuba, Tsukuba, Ibaraki 305, Japan} 
\affiliation{Tufts University, Medford, Massachusetts 02155} 
\affiliation{Waseda University, Tokyo 169, Japan} 
\affiliation{Wayne State University, Detroit, Michigan  48201} 
\affiliation{University of Wisconsin, Madison, Wisconsin 53706} 
\affiliation{Yale University, New Haven, Connecticut 06520} 
\author{T.~Aaltonen}
\affiliation{Division of High Energy Physics, Department of Physics, University of Helsinki and Helsinki Institute of Physics, FIN-00014, Helsinki, Finland}
\author{J.~Adelman}
\affiliation{Enrico Fermi Institute, University of Chicago, Chicago, Illinois 60637}
\author{T.~Akimoto}
\affiliation{University of Tsukuba, Tsukuba, Ibaraki 305, Japan}
\author{B.~\'{A}lvarez~Gonz\'{a}lez$^r$}
\affiliation{Instituto de Fisica de Cantabria, CSIC-University of Cantabria, 39005 Santander, Spain}
\author{S.~Amerio$^w$}
\affiliation{Istituto Nazionale di Fisica Nucleare, Sezione di Padova-Trento, $^w$University of Padova, I-35131 Padova, Italy} 

\author{D.~Amidei}
\affiliation{University of Michigan, Ann Arbor, Michigan 48109}
\author{A.~Anastassov}
\affiliation{Northwestern University, Evanston, Illinois  60208}
\author{A.~Annovi}
\affiliation{Laboratori Nazionali di Frascati, Istituto Nazionale di Fisica Nucleare, I-00044 Frascati, Italy}
\author{J.~Antos}
\affiliation{Comenius University, 842 48 Bratislava, Slovakia; Institute of Experimental Physics, 040 01 Kosice, Slovakia}
\author{G.~Apollinari}
\affiliation{Fermi National Accelerator Laboratory, Batavia, Illinois 60510}
% ...

\collaboration{CDF Collaboration\footnote{With visitors from $^a$University of Massachusetts Amherst, Amherst, Massachusetts 01003,
$^b$Universiteit Antwerpen, B-2610 Antwerp, Belgium, 
$^c$University of Bristol, Bristol BS8 1TL, United Kingdom,
$^d$Chinese Academy of Sciences, Beijing 100864, China, 
$^e$Istituto Nazionale di Fisica Nucleare, Sezione di Cagliari, 09042 Monserrato (Cagliari), Italy,
$^f$University of California Irvine, Irvine, CA  92697, 
$^g$University of California Santa Cruz, Santa Cruz, CA  95064, 
$^h$Cornell University, Ithaca, NY  14853, 
$^i$University of Cyprus, Nicosia CY-1678, Cyprus, 
$^j$University College Dublin, Dublin 4, Ireland,
$^k$Royal Society of Edinburgh/Scottish Executive Support Research Fellow,
$^l$University of Edinburgh, Edinburgh EH9 3JZ, United Kingdom, 
$^m$Universidad Iberoamericana, Mexico D.F., Mexico,
$^n$Queen Mary, University of London, London, E1 4NS, England,
$^o$University of Manchester, Manchester M13 9PL, England, 
$^p$Nagasaki Institute of Applied Science, Nagasaki, Japan, 
$^q$University of Notre Dame, Notre Dame, IN 46556,
$^r$University de Oviedo, E-33007 Oviedo, Spain, 
$^s$Texas Tech University, Lubbock, TX  79409, 
$^t$IFIC(CSIC-Universitat de Valencia), 46071 Valencia, Spain,
$^u$University of Virginia, Charlottesville, VA  22904,
$^{cc}$On leave from J.~Stefan Institute, Ljubljana, Slovenia, 
}}
\noaffiliation
 
\pagestyle{empty}





\date{\today}

\begin{abstract}
 A search for a narrow diphoton mass resonance is presented based on
 data from 3.0~fb$^{-1}$ of integrated luminosity from $p\bar p$
 collisions at $\sqrt{s}$ = 1.96 TeV collected by the CDF experiment.
 No evidence of a resonance in the diphoton mass spectrum is observed,
 and upper limits are set on the cross section times branching
 fraction of the resonant state as a function of Higgs boson mass.
 The resulting limits exclude Higgs bosons with masses below
 106~\gevcc\ at a 95\% Bayesian credibility level (C.L.) for one
 fermiophobic benchmark model.
\end{abstract}


\maketitle


  The standard model (SM) of particle physics has proven to be an
extremely robust theory through its accurate predictions of many
experimental results obtained over the last few decades.  Although the
Higgs mechanism~\cite{Higgs:1964pj} was proposed in the 1960's, the
particle it predicts, the Higgs boson ($h$), has yet to be observed in
nature.

  The SM prediction for the {$h \rightarrow \gamma \gamma$} branching
  fraction is extremely small (reaching a maximal value of only about
  0.2\% at a Higgs boson mass (\Mh)$~\sim
  120$~\gevcc)~\cite{Djouadi:1997yw}; however, in ``fermiophobic''
  models, where the coupling of the Higgs boson to
  fermions is suppressed, the diphoton decay 
can be greatly enhanced.  
This phenomenon has been
  shown to arise in a variety of extensions to the
  SM~\cite{Haber:1978jt,Gunion:1989ci,Basdevant:1992nb,Barger:1992ty,Akeroyd:1995hg},
  and the resulting collider phenomenology has been
  described~\cite{Dobrescu:1999gv,Matchev1,Mrenna}.  For this
  fermiophobic case, the diphoton final state dominates at low Higgs
  boson masses and is therefore the preferred search channel.

\begin{figure}[h]

\includegraphics[width=0.45\linewidth]{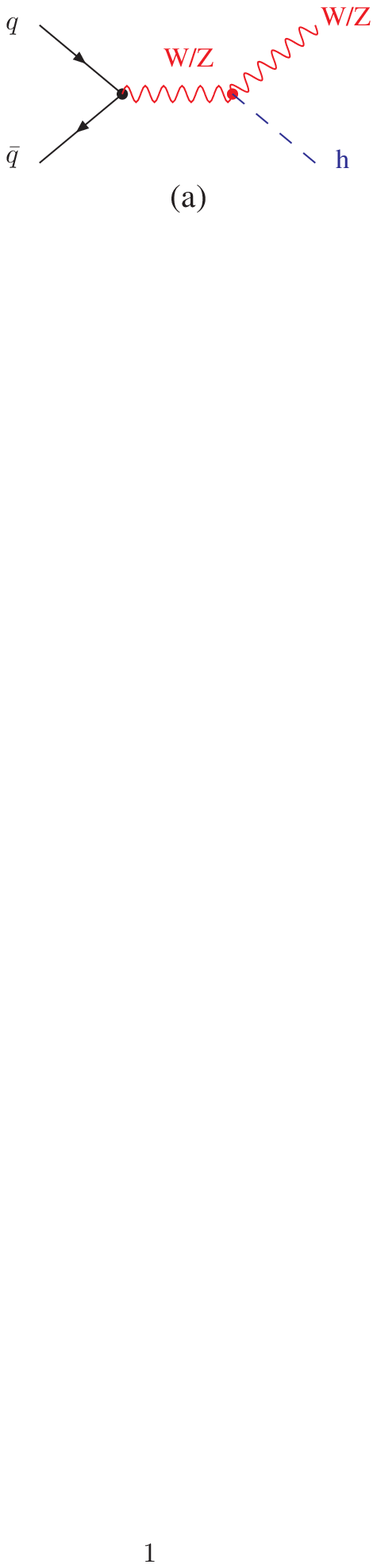}
\includegraphics[width=0.45\linewidth]{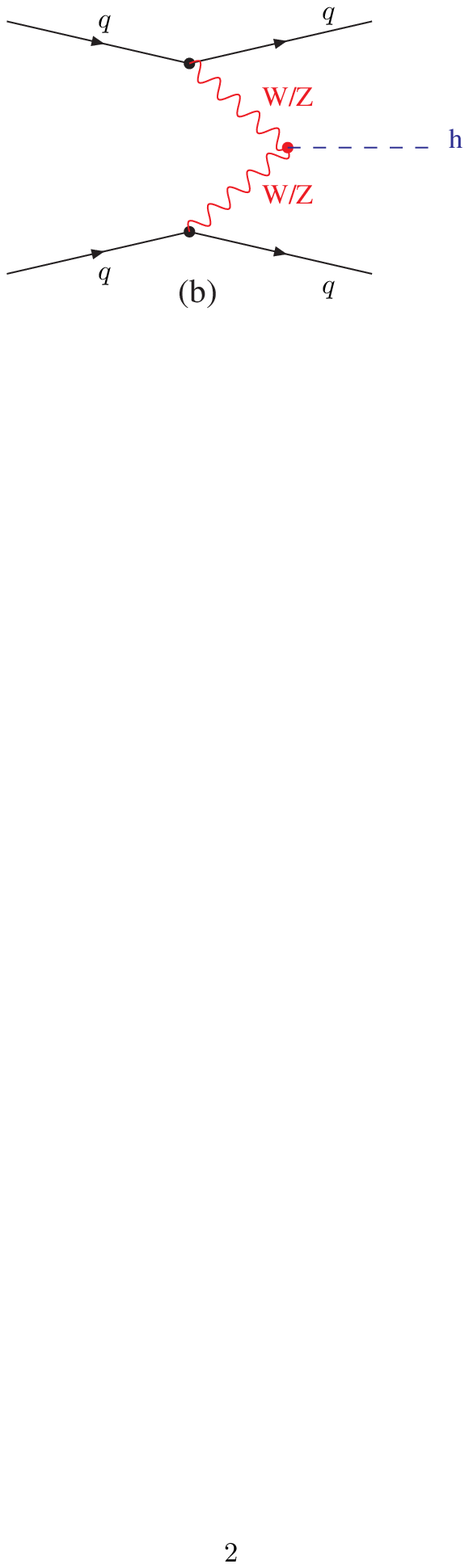}
\caption{The dominant production diagrams for the benchmark
fermiophobic Higgs boson model: associated production with a vector boson (a), and vector boson fusion (b).}
\label{fig:Production}
\end{figure}

 A benchmark fermiophobic model is considered in which the Higgs boson
 does not couple to fermions, yet retains its SM couplings
 to bosons.  In this model, the fermiophobic Higgs boson production is
 dominated by two processes: associated production (shown in
 Fig.~\ref{fig:Production}(a)), and vector boson fusion (abbr. VBF, shown in
 Fig.~\ref{fig:Production}(b)).

Each of the four
experiments~\cite{Heister:2002ub,Abdallah:2003xf,Achard:2003jb,Abbiendi:2002yc}
at the LEP electron-positron collider at CERN place 95\% C.L. lower
limits on the fermiophobic Higgs boson mass (the most stringent being
105.5~\gevcc), while a combination of these results obtains a 95\%
C.L. limit of 109.7 \gevcc~\cite{Rosca:2002me}.  The CDF and D0
experiments at the Tevatron also searched for a fermiophobic Higgs
boson \cite{Affolder:2001hx,Abbott:1998vv}.  Most recently, the D0
experiment set limits on the production cross section of the
fermiophobic Higgs boson with 1.1 \fb\ of data, resulting in a 95\%
C.L. lower limit on \Mh\ of 100 \gevcc~\cite{hf_D0}.  In this
Letter, we search the diphoton mass distribution from the Collider Detector at Fermilab (CDF) for a narrow resonance that could reveal the
presence of a fermiophobic Higgs boson. 

We use the CDF II detector~\cite{Acosta:2004yw,geometry} to identify (ID) photon candidate
events produced in $p\bar p$ collisions at $\sqrt{s}$ = 1.96 TeV.  
The innermost detector component is the silicon vertex tracker~\cite{Sill:2000zz} which is surrounded by an open-cell drift
chamber (COT,~\cite{Affolder:2003ep}).  Both sample the
trajectories of charged particles and determine their momentum as they
curve in the presence of a 1.4 T axial magnetic field.  

Particles that pass through the COT reach the
electromagnetic and hadronic
calorimeters~\cite{ccal,wha,Albrow:2001jw}, which are divided into two
regions: central ($|\eta|<1.1$) and forward or ``plug''
($1.1<|\eta|<3.6$).  At the approximate electromagnetic shower
maximum, the calorimeters contain fine-grained
detectors~\cite{Apollinari:1998bg} that measure the shower shape and
centroid position in the two dimensions transverse to the shower
development.  The calorimeters are surrounded by a system of muon
chambers~\cite{Ascoli:1987av}.

 Three levels of real-time event selection (trigger) systems are used
to filter events.  The trigger paths used here require two clusters of
deposited energy in the electromagnetic calorimeter.  One path
requires that both clusters have a transverse energy \et\
$>12$~\gev~\cite{geometry} and be isolated from other energy clusters
in the calorimeter~\cite{Wynne:2009}.  A second trigger has a cluster
transverse energy requirement of \et\ $>18$~\gev\ without the
requirement of cluster isolation.  By combining these two trigger paths,
virtually all of the identifiable diphoton events are recorded.

The analysis is divided into two independent subsamples according to
the position of the photons: the first requires that both photons be
located within the fiducial region of the central electromagnetic
calorimeter ($|\eta|< 1.05$), and the second requires that one photon
be located in this region and the other in the plug calorimeter
($1.2<|\eta|<2.8$).  The former will be referred to as central-central
(CC) events, and the latter as central-plug (CP) events~\cite{Z_pole}.
The data were recorded between February of 2002 and April of 2008,
corresponding to an integrated luminosity of 3.0 \fb\ for CC and 2.9
\fb\ for CP events. 

A series of baseline selection criteria helps to remove background
events and to ID high-energy photon candidates for the
analysis.  Individual photons are required to have \et\ $>15$~\gev,
while the diphoton pair is required to have mass \Mgg\ $>30$~\gevcc.
Photons are required to pass CDF standard photon ID requirements
including the following~\cite{Wynne:2009,RS_PRL}: transverse shower
profiles consistent with single photon expectation from test beam
studies~\cite{Abe:1993qb}, additional transverse energy in the
calorimeter in a cone of angular radius $R = \sqrt{(\Delta \phi)^{2}
+ (\Delta
\eta)^{2}} = 0.4$~\cite{geometry} around the photon candidate be
less than 2~GeV, and the scalar sum of the $\ptm$ of the tracks in the
same cone be less than 2~GeV$/c$.  Central photons must also be
isolated in the shower maximum detector.

   The above selection criteria define an inclusive diphoton sample.
   However, the fermiophobic Higgs boson is only produced at a
   non-negligible rate in association with a $W$ or $Z$ boson or via
   the VBF process.  In order to improve sensitivity, the event
   selection was further extended to take advantage of the final state
   features present in these production modes. Associated production
   dominates the production process (for \Mh\ = 100~\gevcc\, its rate
   is about four times larger than VBF), so the optimization was
   carried out on the basis of the associated production process
   alone. A selection based on the following observables
   was optimized: diphoton transverse momentum (\ptgg), transverse
   momentum of the second highest \pt\ jet ($p_T^{~j2}$) for hadronic
   decays of $W/Z$, and missing transverse energy (\met) or transverse
   momentum of an isolated track ($p_T^{~iso}$) for leptonic decays of
   $W/Z$.  

  For the optimization study, a Bayesian method with a flat prior
  probability was used to estimate the expected limits based on signal
  and background event expectations in a $10$~\gevcc\ mass window
  centered at 100~\gevcc.  The diphoton background is composed of SM
  diphoton events ($\sim$ 25~\%) and events in which either one
  or both photon candidates are actually quark or gluon jets which
  were misidentified as photons ($\sim$ 75~\%).  Higgs boson events
  with only the diphoton decay mode and SM diphoton events were
  generated using the \py\ 6.2~\cite{PYTHIA} Monte Carlo event
  generator and a parametrized response of the CDF II
  detector~\cite{GEANT3,GFLASH}.  All \py\ samples were made with
  CTEQ5L~\cite{CTEQ5L} parton distribution functions, where the \py\
  underlying event model is tuned to CDF jet data~\cite{TUNEA}.  The
  background component arising from jets misidentified as photons was
  estimated using photon identification control regions from data.
  The control regions do not overlap with the signal region, as the
  events in the control region are required to fail at least one of
  the standard electromagnetic energy fraction or isolation
  requirements, yet pass a looser set of these requirements.  

 The optimization shows that a requirement of \ptgg\ $>75$~\gevc\ is approximately as
 sensitive as any combination of the other selection criteria.  With
 this requirement on \ptgg, roughly 30\% of the signal remains
 (slightly varying with \Mh) while more than 99.5\% of the background is
 removed.  Although the cut was optimized based on associated
 production, VBF also has a higher average \ptgg\ than the background
 processes and is included in the analysis with the same selection.

  The detector acceptance for signal events is calculated using
the \py\ event generator samples described above.  Since a pure sample of
reconstructed photons is not available in the data, corrections to the
photon identification efficiencies due to imperfections in the
detector simulation are derived using electrons from $Z$ boson decays.
This is justified since the energy deposition in the EM calorimeter by
electrons and photons is almost indistinguishable.  The electrons are
selected with a slightly modified version of the photon ID
requirements to allow the presence of a high \pt\ track.  A
correction factor to the ID efficiency of the simulation of 0.97
(0.94) is derived for central (plug) photons by comparing ID
efficiencies from the detector simulation with the ID efficiencies
measured in data.

The largest systematic uncertainties on the expected number of Higgs
boson events arise from the luminosity measurement (6\%), varying the
parameters controlling the amount of initial and final state radiation
from the parton shower model of \py\ ($4\%$)~\cite{ISR}, and the \py\
modeling of the shape of
the \ptgg\ distribution for the signal ($4\%$).  The latter
uncertainty was obtained by studying the effect on the acceptance from
the differences in the shape of the \ptgg\ distribution from
leading-order, next-to-leading-order, and \py\
predictions~\cite{MrennaQCD}.  Other systematic uncertainties were
also considered due to uncertainties in photon ID efficiency, the
electromagnetic energy scale, and parton distribution
functions~\cite{CTEQ6.1M,Weights}.  The signal acceptances are included
in Table~\ref{tab:limits} and they have a relative uncertainty
of 8\% (9\%) for CC (CP).

\begin{figure}[h]
\begin{center}
\includegraphics[width=1.1\linewidth]{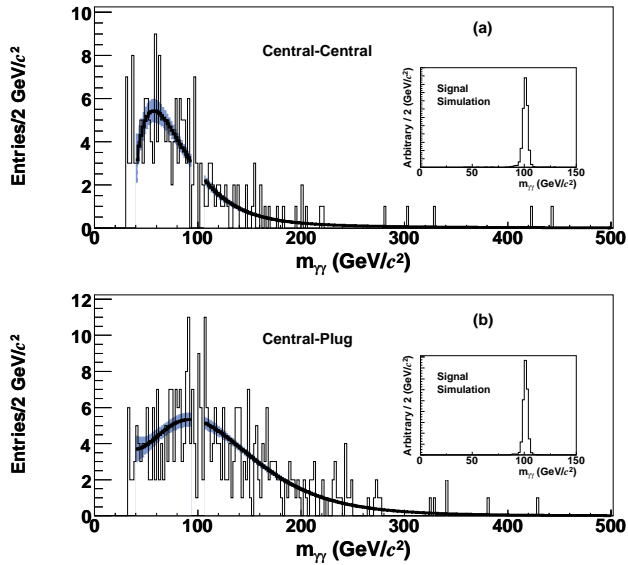}
\end{center}
\caption{The invariant mass distribution of central-central (a) and
central-plug (b) photon pairs after the requirement of \ptgg\
$>75$~\gevc, shown with the fit to the data for the hypothesis of a
\Mh\ of 100~\gevcc.  The gap in the fit centered at 100~\gevcc\
represents the signal region for this mass point that was excluded
from the fit.  The error bands show the statistical uncertainty in the fit.  The expected shape of the signal from simulation is shown in the inserts.}  
\label{fig:Error_Bands}
\end{figure}

  The decay of a Higgs boson into a diphoton pair appears as
   a very narrow peak in the invariant mass distribution of these two
   photons.  The
   diphoton mass resolution as determined from simulation is better
   than 3\%\ for the Higgs boson mass region studied here and is
   limited by the energy resolution of the electromagnetic calorimeters.  The
  simulated resolution was compared with data using
  electrons from $Z$ boson decays and is found to model the detector
  well. The diphoton invariant mass distribution shown in
  Fig.~\ref{fig:Error_Bands} could be sensitive to a resonance in the
  context of the fermiophobic benchmark model (see the inserts in
  Fig.~\ref{fig:Error_Bands} for the signal shape expected from
  simulation). No evidence of such a resonance is apparent in the
  data.  As a background prediction to be used for setting limits on a
  diphoton resonance, the data is fit to a sum of two exponentials
  multiplied by a fractional degree polynomial, where the degree of
  one term is a parameter of the fit.  The fit, excluding a 10~\gevcc\
  mass window centered at each test \Mh, is performed for each \Mh\
  hypothesis (10-\gevcc\ steps from 70 to 150~\gevcc).  The fit for a \Mh\
  of 100~\gevcc\ is shown in Fig.~\ref{fig:Error_Bands}.

After the background fit for each mass hypothesis has been determined,
the presence or absence of a Higgs boson signal is ascertained on the basis
of a binned likelihood method incorporating the simulated signal shape
and the systematic uncertainties. We calculate a Bayesian C.L. limit
for each mass hypothesis based on the combined binned likelihood of
the mass distributions for the CC and CP samples.  A posterior density
in $\sigma\times$\BRhgg\ is obtained by multiplying this likelihood by Gaussian prior densities for the background normalizations and systematic uncertainties leaving
the production cross section ($\sigma\times$\BRhgg) with a uniform
prior density.  A 95\% C.L. limit is then determined such that 95\% of
the posterior density for $\sigma\times$\BRhgg\ falls below the
limit~\cite{Amsler:2008zz}.

\begin{figure}[h]
\begin{center}
\includegraphics[width=1.1\linewidth]{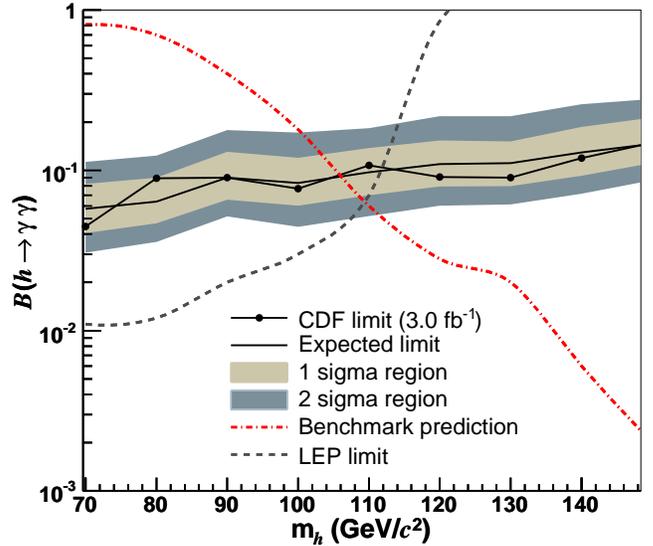}
\end{center}
\caption{The 95\% C.L. upper limit on the branching
fraction for the fermiophobic Higgs boson decay to diphotons, as a
function of \Mh.  The shaded regions represent the one and two sigma
probability of fluctuations of the observed limit away from the expected limit based on the distribution of possible experimental outcomes.  For reference, the 95\% C.L. limits from LEP are also included.}  
\label{fig:Limits}
\end{figure}

The results of the limit calculation are included in
Table~\ref{tab:limits} and displayed graphically in
Fig.~\ref{fig:Limits}.  The SM cross sections assumed in
the benchmark fermiophobic model are used to convert the limits on
$\sigma\times$\BRhgg\ into limits on \BRhgg. The result excludes the
benchmark model predictions (at 95\% C.L.) for \Mh\ of less than 106
\gevcc.     

\begin{table*}
\caption{Observed and expected 95\% C.L. limits on the production
cross section and branching fraction and theory predictions for the
fermiophobic benchmark Higgs boson model. Total signal acceptance is also
included.  Note that the CP channel is not included for the 90~\gevcc\ point~\cite{Z_pole}. \label{tab:limits}} 
\begin{center}
\begin{tabular}{cccccccc}
\hline
\hline
  & & \multicolumn{3}{c}{ $\sigma \times $\BRhgg\ (fb)} & \multicolumn{3}{c}{\BRhgg\ (\%)} \\ \hline
 &  Acceptance (\%) & \multicolumn{2}{c}{Limits} & & \multicolumn{2}{c}{Limits} & \\
$m_{h}$ (\gevcc)                        &
(CC +CP)            &
Expected            &
Observed            &
Theory~\cite{Djouadi:1997yw}        &
Expected            &
Observed            &
Theory           \\
\hline
70   & 4.8 &88.1  &68.2 & 1240   & 5.8   & 4.5  & 81   \\	   
80   & 6.2 &68.3  &95.4 & 749    & 6.4   & 8.9  & 70   \\	   
90   & 4.4 & 70.7 &70.8 & 312    & 9.1   & 9.1 & 40 \\	   
100  & 8.8 &48.3  &44.5 & 104    & 8.3   & 7.7  & 18   \\	   
110  & 10  &41.8  &46.2 & 25.8   & 9.7   & 10.7 & 6.0  \\	   
120  & 11  &36.3  &30.2 & 9.3    & 10.9  & 9.1  & 2.8  \\	   
130  & 12  &27.8  &22.6 & 5.0    & 11.0  & 9.0  & 2.0  \\	   
140  & 13  &26.6  &24.4 & 1.2    & 12.0  & 11.9 & 0.6  \\	    
150  & 15  &23.5  &23.9 & 0.3    & 14.6  & 14.9 & 0.2  \\\hline
\hline
 \end{tabular}
\end{center}
\end{table*}

This Letter presents the results of a search for a narrow resonance in
the diphoton mass spectrum using data comprising 3.0 \fb\ of
integrated luminosity from the CDF II detector at the Tevatron.
Selected events include two photons in the central detector or one
photon in the central and one in the plug detector.  There is no
evidence of a narrow resonance.  Limits are placed on the production
cross section and the branching fraction for the Higgs
boson decay into a photon pair and compared to the predictions of a
benchmark fermiophobic model resulting in a limit on the Higgs boson mass
of \Mh\ $>$ 106 \gevcc\ at the 95\% C.L.
This mass limit is approximately as strong as any previous single
experiment, and the result significantly extends the excluded region of
\BRhgg\ for \Mh\ above 110~\gevcc.

We thank the Fermilab staff and the technical staffs of the
participating institutions for their vital contributions.  This work
was supported by the U.S.  Department of Energy and National Science
Foundation; the Italian Istituto Nazionale di Fisica Nucleare; the
Ministry of Education, Culture, Sports, Science and Technology of
Japan; the Natural Sciences and Engineering Research Council of
Canada; the National Science Council of the Republic of China; the
Swiss National Science Foundation; the A.P.  Sloan Foundation; the
Bundesministerium f\"ur Bildung und Forschung, Germany; the Korean
Science and Engineering Foundation and the Korean Research Foundation;
the Science and Technology Facilities Council and the Royal Society,
UK; the Institut National de Physique Nucleaire et Physique des
Particules/CNRS; the Russian Foundation for Basic Research; the
Ministerio de Ciencia e Innovaci\'{o}n, and Programa
Consolider-Ingenio 2010, Spain; the Slovak R\&D Agency; and the
Academy of Finland.

{
\bibliographystyle{apsrev}
\bibliography{HGammaGammaPRL}

\begin{thebibliography}{41}
\expandafter\ifx\csname natexlab\endcsname\relax\def\natexlab#1{#1}\fi
\expandafter\ifx\csname bibnamefont\endcsname\relax
  \def\bibnamefont#1{#1}\fi
\expandafter\ifx\csname bibfnamefont\endcsname\relax
  \def\bibfnamefont#1{#1}\fi
\expandafter\ifx\csname citenamefont\endcsname\relax
  \def\citenamefont#1{#1}\fi
\expandafter\ifx\csname url\endcsname\relax
  \def\url#1{\texttt{#1}}\fi
\expandafter\ifx\csname urlprefix\endcsname\relax\def\urlprefix{URL }\fi
\providecommand{\bibinfo}[2]{#2}
\providecommand{\eprint}[2][]{\url{#2}}

\bibitem[{\citenamefont{Higgs}(1964)}]{Higgs:1964pj}
\bibinfo{author}{\bibfnamefont{P.~W.} \bibnamefont{Higgs}},
  \bibinfo{journal}{Phys. Rev. Lett.} \textbf{\bibinfo{volume}{13}},
  \bibinfo{pages}{508} (\bibinfo{year}{1964}).

\bibitem[{\citenamefont{Djouadi et~al.}(1998)\citenamefont{Djouadi, Kalinowski,
  and Spira}}]{Djouadi:1997yw}
\bibinfo{author}{\bibfnamefont{A.}~\bibnamefont{Djouadi}},
  \bibinfo{author}{\bibfnamefont{J.}~\bibnamefont{Kalinowski}},
  \bibnamefont{and} \bibinfo{author}{\bibfnamefont{M.}~\bibnamefont{Spira}},
  \bibinfo{journal}{Comput. Phys. Commun.} \textbf{\bibinfo{volume}{108}},
  \bibinfo{pages}{56} (\bibinfo{year}{1998}).

\bibitem[{\citenamefont{Haber et~al.}(1979)\citenamefont{Haber, Kane, and
  Sterling}}]{Haber:1978jt}
\bibinfo{author}{\bibfnamefont{H.~E.} \bibnamefont{Haber}},
  \bibinfo{author}{\bibfnamefont{G.~L.} \bibnamefont{Kane}}, \bibnamefont{and}
  \bibinfo{author}{\bibfnamefont{T.}~\bibnamefont{Sterling}},
  \bibinfo{journal}{Nucl. Phys. B} \textbf{\bibinfo{volume}{161}},
  \bibinfo{pages}{493} (\bibinfo{year}{1979}).

\bibitem[{\citenamefont{Gunion et~al.}(1990)\citenamefont{Gunion, Vega, and
  Wudka}}]{Gunion:1989ci}
\bibinfo{author}{\bibfnamefont{J.~F.} \bibnamefont{Gunion}},
  \bibinfo{author}{\bibfnamefont{R.}~\bibnamefont{Vega}}, \bibnamefont{and}
  \bibinfo{author}{\bibfnamefont{J.}~\bibnamefont{Wudka}},
  \bibinfo{journal}{Phys. Rev. D} \textbf{\bibinfo{volume}{42}},
  \bibinfo{pages}{1673} (\bibinfo{year}{1990}).

\bibitem[{\citenamefont{Basdevant et~al.}(1993)}]{Basdevant:1992nb}
\bibinfo{author}{\bibfnamefont{J.~L.} \bibnamefont{Basdevant}}
  \bibnamefont{et~al.}, \bibinfo{journal}{Phys. Lett. B}
  \textbf{\bibinfo{volume}{313}}, \bibinfo{pages}{402} (\bibinfo{year}{1993}).

\bibitem[{\citenamefont{Barger et~al.}(1992)}]{Barger:1992ty}
\bibinfo{author}{\bibfnamefont{V.}~\bibnamefont{Barger}} \bibnamefont{et~al.}
  (\bibinfo{year}{1992}), \eprint{arXiv:hep-ph/9211234}.

\bibitem[{\citenamefont{Akeroyd}(1996)}]{Akeroyd:1995hg}
\bibinfo{author}{\bibfnamefont{A.~G.} \bibnamefont{Akeroyd}},
  \bibinfo{journal}{Phys. Lett. B} \textbf{\bibinfo{volume}{368}},
  \bibinfo{pages}{89} (\bibinfo{year}{1996}).

\bibitem[{\citenamefont{Dobrescu}(2001)}]{Dobrescu:1999gv}
\bibinfo{author}{\bibfnamefont{B.~A.} \bibnamefont{Dobrescu}},
  \bibinfo{journal}{Phys. Rev.} \textbf{\bibinfo{volume}{D63}},
  \bibinfo{pages}{015004} (\bibinfo{year}{2001}).

\bibitem[{\citenamefont{Landsberg and Matchev}(2000)}]{Matchev1}
\bibinfo{author}{\bibfnamefont{G.~L.} \bibnamefont{Landsberg}}
  \bibnamefont{and} \bibinfo{author}{\bibfnamefont{K.~T.}
  \bibnamefont{Matchev}}, \bibinfo{journal}{Phys. Rev. D}
  \textbf{\bibinfo{volume}{62}}, \bibinfo{pages}{035004}
  (\bibinfo{year}{2000}).

\bibitem[{\citenamefont{Mrenna and Wells}(2001)}]{Mrenna}
\bibinfo{author}{\bibfnamefont{S.}~\bibnamefont{Mrenna}} \bibnamefont{and}
  \bibinfo{author}{\bibfnamefont{J.~D.} \bibnamefont{Wells}},
  \bibinfo{journal}{Phys. Rev. D} \textbf{\bibinfo{volume}{63}},
  \bibinfo{pages}{015006} (\bibinfo{year}{2001}).

\bibitem[{\citenamefont{Heister et~al.}(2002)}]{Heister:2002ub}
\bibinfo{author}{\bibfnamefont{A.}~\bibnamefont{Heister}} \bibnamefont{et~al.}
  (\bibinfo{collaboration}{ALEPH Collaboration}), \bibinfo{journal}{Phys. Lett.
  B} \textbf{\bibinfo{volume}{544}}, \bibinfo{pages}{16}
  (\bibinfo{year}{2002}).

\bibitem[{\citenamefont{Abdallah et~al.}(2004)}]{Abdallah:2003xf}
\bibinfo{author}{\bibfnamefont{J.}~\bibnamefont{Abdallah}} \bibnamefont{et~al.}
  (\bibinfo{collaboration}{DELPHI Collaboration}), \bibinfo{journal}{Eur. Phys.
  J. C} \textbf{\bibinfo{volume}{35}}, \bibinfo{pages}{313}
  (\bibinfo{year}{2004}).

\bibitem[{\citenamefont{Achard et~al.}(2003)}]{Achard:2003jb}
\bibinfo{author}{\bibfnamefont{P.}~\bibnamefont{Achard}} \bibnamefont{et~al.}
  (\bibinfo{collaboration}{L3 Collaboration}), \bibinfo{journal}{Phys. Lett. B}
  \textbf{\bibinfo{volume}{568}}, \bibinfo{pages}{191} (\bibinfo{year}{2003}).

\bibitem[{\citenamefont{Abbiendi et~al.}(2002)}]{Abbiendi:2002yc}
\bibinfo{author}{\bibfnamefont{G.}~\bibnamefont{Abbiendi}} \bibnamefont{et~al.}
  (\bibinfo{collaboration}{OPAL Collaboration}), \bibinfo{journal}{Phys. Lett.
  B} \textbf{\bibinfo{volume}{544}}, \bibinfo{pages}{44}
  (\bibinfo{year}{2002}).

\bibitem[{\citenamefont{Rosca}(2002)}]{Rosca:2002me}
\bibinfo{author}{\bibfnamefont{A.}~\bibnamefont{Rosca}}
  (\bibinfo{collaboration}{LEP}) (\bibinfo{year}{2002}),
  \eprint{arXiv:hep-ex/0212038}.

\bibitem[{\citenamefont{Affolder et~al.}(2001)}]{Affolder:2001hx}
\bibinfo{author}{\bibfnamefont{A.~A.} \bibnamefont{Affolder}}
  \bibnamefont{et~al.} (\bibinfo{collaboration}{CDF Collaboration}),
  \bibinfo{journal}{Phys. Rev. D} \textbf{\bibinfo{volume}{64}},
  \bibinfo{pages}{092002} (\bibinfo{year}{2001}).

\bibitem[{\citenamefont{Abbott et~al.}(1999)}]{Abbott:1998vv}
\bibinfo{author}{\bibfnamefont{B.}~\bibnamefont{Abbott}} \bibnamefont{et~al.}
  (\bibinfo{collaboration}{D0 Collaboration}), \bibinfo{journal}{Phys. Rev.
  Lett.} \textbf{\bibinfo{volume}{82}}, \bibinfo{pages}{2244}
  (\bibinfo{year}{1999}).

\bibitem[{\citenamefont{Abazov et~al.}(2008)}]{hf_D0}
\bibinfo{author}{\bibfnamefont{V.}~\bibnamefont{Abazov}} \bibnamefont{et~al.}
  (\bibinfo{collaboration}{D0 Collaboration}), \bibinfo{journal}{Phys. Rev.
  Lett.} \textbf{\bibinfo{volume}{101}}, \bibinfo{pages}{051801}
  (\bibinfo{year}{2008}).

\bibitem[{\citenamefont{Acosta et~al.}(2005)}]{Acosta:2004yw}
\bibinfo{author}{\bibfnamefont{D.~E.} \bibnamefont{Acosta}}
  \bibnamefont{et~al.} (\bibinfo{collaboration}{CDF Collaboration}),
  \bibinfo{journal}{Phys. Rev. D} \textbf{\bibinfo{volume}{71}},
  \bibinfo{pages}{032001} (\bibinfo{year}{2005}).

\bibitem[{geo()}]{geometry}
\bibinfo{note}{We use a cylindrical coordinate system with its origin in the
  center of the detector, where $\theta$ and $\phi$ are the polar and azimuthal
  angles, respectively, and pseudorapidity is $\eta= - \rm ln$ $\rm
  tan(\theta/2)$. The missing $E_T$ ($\METVEC$) is defined by $\METVEC =
  -\sum_iE_T^i{\hat n}_i, i={\rm calorimeter~tower~number}$, where ${\hat n}_i$
  is a unit vector perpendicular to the beam axis and pointing at the $i^{th}$
  calorimeter tower. $\METVEC$ is corrected for high-energy muons and also jet
  energy corrections. We define $\MET=|\METVEC|$. The transverse momentum $p_T$
  is defined to be $p\sin\theta$.}

\bibitem[{\citenamefont{Sill}(2000)}]{Sill:2000zz}
\bibinfo{author}{\bibfnamefont{A.}~\bibnamefont{Sill}}
  (\bibinfo{collaboration}{CDF Collaboration}), \bibinfo{journal}{Nucl.
  Instrum. Methods} \textbf{\bibinfo{volume}{447}}, \bibinfo{pages}{1}
  (\bibinfo{year}{2000}).

\bibitem[{\citenamefont{Affolder et~al.}(2004)}]{Affolder:2003ep}
\bibinfo{author}{\bibfnamefont{A.~A.} \bibnamefont{Affolder}}
  \bibnamefont{et~al.} (\bibinfo{collaboration}{CDF Collaboration}),
  \bibinfo{journal}{Nucl. Instrum. Methods} \textbf{\bibinfo{volume}{526}},
  \bibinfo{pages}{249} (\bibinfo{year}{2004}).

\bibitem[{\citenamefont{Balka et~al.}(1988)}]{ccal}
\bibinfo{author}{\bibfnamefont{L.}~\bibnamefont{Balka}} \bibnamefont{et~al.}
  (\bibinfo{collaboration}{CDF Collaboration}), \bibinfo{journal}{Nucl.
  Instrum. Methods} \textbf{\bibinfo{volume}{267}}, \bibinfo{pages}{272}
  (\bibinfo{year}{1988}).

\bibitem[{\citenamefont{Bertolucci et~al.}(1988)}]{wha}
\bibinfo{author}{\bibfnamefont{S.}~\bibnamefont{Bertolucci}}
  \bibnamefont{et~al.} (\bibinfo{collaboration}{CDF Collaboration}),
  \bibinfo{journal}{Nucl. Instrum. Methods} \textbf{\bibinfo{volume}{267}},
  \bibinfo{pages}{301} (\bibinfo{year}{1988}).

\bibitem[{\citenamefont{Albrow et~al.}(2002)}]{Albrow:2001jw}
\bibinfo{author}{\bibfnamefont{M.~G.} \bibnamefont{Albrow}}
  \bibnamefont{et~al.} (\bibinfo{collaboration}{CDF Collaboration}),
  \bibinfo{journal}{Nucl. Instrum. Methods} \textbf{\bibinfo{volume}{480}},
  \bibinfo{pages}{524} (\bibinfo{year}{2002}).

\bibitem[{\citenamefont{Apollinari et~al.}(1998)}]{Apollinari:1998bg}
\bibinfo{author}{\bibfnamefont{G.}~\bibnamefont{Apollinari}}
  \bibnamefont{et~al.}, \bibinfo{journal}{Nucl. Instrum. Methods}
  \textbf{\bibinfo{volume}{412}}, \bibinfo{pages}{515} (\bibinfo{year}{1998}).

\bibitem[{\citenamefont{Ascoli et~al.}(1988)}]{Ascoli:1987av}
\bibinfo{author}{\bibfnamefont{G.}~\bibnamefont{Ascoli}} \bibnamefont{et~al.},
  \bibinfo{journal}{Nucl. Instrum. Methods} \textbf{\bibinfo{volume}{268}},
  \bibinfo{pages}{33} (\bibinfo{year}{1988}).

\bibitem[{\citenamefont{Wynne}()}]{Wynne:2009}
\bibinfo{author}{\bibfnamefont{S.}~\bibnamefont{Wynne}}, \bibinfo{note}{{P}h.D.
  thesis, Liverpool University, {F}ERMILAB-THESIS-2007-17}.

\bibitem[{Z_p()}]{Z_pole}
\bibinfo{note}{The CP channel is not used for the 90~\gevcc\ Higgs mass
  hypothesis due to reduced sensitivity caused by contamination from the $Z$
  boson mass peak.}

\bibitem[{\citenamefont{Aaltonen et~al.}(2007)}]{RS_PRL}
\bibinfo{author}{\bibfnamefont{T.}~\bibnamefont{Aaltonen}} \bibnamefont{et~al.}
  (\bibinfo{collaboration}{CDF Collaboration}), \bibinfo{journal}{Phys. Rev.
  Lett.} \textbf{\bibinfo{volume}{99}}, \bibinfo{pages}{171801}
  (\bibinfo{year}{2007}).

\bibitem[{\citenamefont{Abe et~al.}(1993)}]{Abe:1993qb}
\bibinfo{author}{\bibfnamefont{F.}~\bibnamefont{Abe}} \bibnamefont{et~al.}
  (\bibinfo{collaboration}{CDF Collaboration}), \bibinfo{journal}{Phys. Rev. D}
  \textbf{\bibinfo{volume}{48}}, \bibinfo{pages}{2998} (\bibinfo{year}{1993}).

\bibitem[{\citenamefont{Sjostrand et~al.}(2001)}]{PYTHIA}
\bibinfo{author}{\bibfnamefont{T.}~\bibnamefont{Sjostrand}}
  \bibnamefont{et~al.}, \bibinfo{journal}{Comput. Phys. Commun.}
  \textbf{\bibinfo{volume}{135}}, \bibinfo{pages}{238} (\bibinfo{year}{2001}).

\bibitem[{\citenamefont{Brun et~al.}(1987)}]{GEANT3}
\bibinfo{author}{\bibfnamefont{R.}~\bibnamefont{Brun}} \bibnamefont{et~al.}
  (\bibinfo{year}{1987}), \bibinfo{note}{unpublished, {C}ERN-DD/EE/84-1}.

\bibitem[{\citenamefont{Grindhammer et~al.}(1990)\citenamefont{Grindhammer,
  Rudowicz, and Peters}}]{GFLASH}
\bibinfo{author}{\bibfnamefont{G.}~\bibnamefont{Grindhammer}},
  \bibinfo{author}{\bibfnamefont{M.}~\bibnamefont{Rudowicz}}, \bibnamefont{and}
  \bibinfo{author}{\bibfnamefont{S.}~\bibnamefont{Peters}},
  \bibinfo{journal}{Nucl. Instrum. Methods} \textbf{\bibinfo{volume}{290}},
  \bibinfo{pages}{469} (\bibinfo{year}{1990}).

\bibitem[{\citenamefont{Lai et~al.}(2000)}]{CTEQ5L}
\bibinfo{author}{\bibfnamefont{H.~L.} \bibnamefont{Lai}} \bibnamefont{et~al.}
  (\bibinfo{collaboration}{CTEQ Collaboration}), \bibinfo{journal}{Eur. Phys.
  J.} \textbf{\bibinfo{volume}{C12}}, \bibinfo{pages}{375}
  (\bibinfo{year}{2000}).

\bibitem[{\citenamefont{Field and Group}(2005)}]{TUNEA}
\bibinfo{author}{\bibfnamefont{R.}~\bibnamefont{Field}} \bibnamefont{and}
  \bibinfo{author}{\bibfnamefont{R.~C.} \bibnamefont{Group}}
  (\bibinfo{collaboration}{CDF Collaboration}) (\bibinfo{year}{2005}),
  \eprint{ArXiv:hep-ph/0510198}.

\bibitem[{ISR()}]{ISR}
\bibinfo{note}{We constrain the rate of initial state radiation using Drell-Yan
  events in data.}

\bibitem[{\citenamefont{Mrenna and Yuan}(1998)}]{MrennaQCD}
\bibinfo{author}{\bibfnamefont{S.}~\bibnamefont{Mrenna}} \bibnamefont{and}
  \bibinfo{author}{\bibfnamefont{C.~P.} \bibnamefont{Yuan}},
  \bibinfo{journal}{Phys. Lett. B} \textbf{\bibinfo{volume}{416}},
  \bibinfo{pages}{200} (\bibinfo{year}{1998}).

\bibitem[{\citenamefont{Stump et~al.}(2003)}]{CTEQ6.1M}
\bibinfo{author}{\bibfnamefont{D.}~\bibnamefont{Stump}} \bibnamefont{et~al.},
  \bibinfo{journal}{J. High Energy Phys.} \textbf{\bibinfo{volume}{0410}},
  \bibinfo{pages}{046} (\bibinfo{year}{2003}).

\bibitem[{\citenamefont{Bourilkov et~al.}(2006)\citenamefont{Bourilkov, Group,
  and Whalley}}]{Weights}
\bibinfo{author}{\bibfnamefont{D.}~\bibnamefont{Bourilkov}},
  \bibinfo{author}{\bibfnamefont{R.~C.} \bibnamefont{Group}}, \bibnamefont{and}
  \bibinfo{author}{\bibfnamefont{M.~R.} \bibnamefont{Whalley}}
  (\bibinfo{year}{2006}), \eprint{arXiv:hep-ph/0605240}.

\bibitem[{\citenamefont{Amsler et~al.}(2008)}]{Amsler:2008zz}
\bibinfo{author}{\bibfnamefont{C.}~\bibnamefont{Amsler}} \bibnamefont{et~al.}
  (\bibinfo{collaboration}{Particle Data Group}), \bibinfo{journal}{Phys. Lett.
  B} \textbf{\bibinfo{volume}{667}}, \bibinfo{pages}{1} (\bibinfo{year}{2008}).

\end{thebibliography}
}



\end{document}